\documentclass[letterpaper, 10 pt, conference]{ieeeconf}  

\IEEEoverridecommandlockouts                              
\usepackage{cite}
\usepackage{amsmath,amssymb,amsfonts}
\usepackage{graphicx}
\usepackage{textcomp}
\usepackage{xcolor}
\usepackage[utf8]{inputenc}
\usepackage[T1]{fontenc}
\usepackage{mathtools}
\usepackage{algorithm}
\usepackage{algorithmic}
\usepackage{bm}
\usepackage{float}
\usepackage{mathrsfs}

\usepackage{relsize} 
\usepackage{hyperref}
\usepackage{amsthm}
\newenvironment{compactproof}{%
  \begin{proof}\begingroup
    \setlength{\abovedisplayskip}{2pt}
    \setlength{\belowdisplayskip}{2pt}
    \setlength{\abovedisplayshortskip}{2pt}
    \setlength{\belowdisplayshortskip}{2pt}
    \linespread{0.8}\selectfont 
}{%
    \endgroup\end{proof}
}

\newenvironment{compactproposition}{\begin{proposition}
  \setlength{\abovedisplayskip}{4pt}
  \setlength{\belowdisplayskip}{4pt}
  \setlength{\abovedisplayshortskip}{2pt}
  \setlength{\belowdisplayshortskip}{2pt}
  \setlength{\arraycolsep}{2pt}
  }{\end{proposition}}

\usepackage{etoolbox}

\newtheorem{theorem}{Theorem}

\theoremstyle{remark}

\newtheorem{definition}{Definition}
\newtheorem{proposition}{Proposition}

\usepackage{svg}
\DeclareMathAlphabet{\mathcal}{OMS}{cmsy}{m}{n}
\usepackage{tikz}
\usetikzlibrary{calc}

\def\BibTeX{{\rm B\kern-.05em{\sc i\kern-.025em b}\kern-.08em
    T\kern-.1667em\lower.7ex\hbox{E}\kern-.125emX}}

\begin{document}

\title{Data-Driven Reachability Analysis for Piecewise Affine Systems
  \thanks{%
    Peng Xie and Amr Alanwar are with the TUM School of Computation, 
    Information and Technology, Department of Computer Engineering, 
    Technical University of Munich, 74076 Heilbronn, Germany.\quad 
    \texttt{(e-mail: p.xie@tum.de, alanwar@tum.de)}\\[4pt]
    Johannes Betz is with the Professorship of Autonomous Vehicle Systems, TUM School of Engineering and Design, Technical University Munich, 85748 Garching, Germany; Munich Institute of Robotics and Machine Intelligence (MIRMI).\quad 
    \texttt{(e-mail: johannes.betz@tum.de)}\\[4pt]
    Davide M.~Raimondo is with the School of Automatic Control, 
    Department of Engineering and Architecture, 
    University of Trieste, Via Valerio 10, 34127 Trieste, Italy.\quad 
    \texttt{(e-mail: DAVIDEMARTINO.RAIMONDO@dia.units.it)}%
  }%
}
\author{Peng Xie, Johannes Betz, Davide M. Raimondo, Amr Alanwar}
\maketitle
\begin{abstract}
Hybrid systems play a crucial role in modeling real-world applications where discrete and continuous dynamics interact, including autonomous vehicles, power systems, and traffic networks. Safety verification for these systems requires determining whether system states can enter unsafe regions under given initial conditions and uncertainties—a question directly addressed by reachability analysis. However, hybrid systems present unique difficulties because their state space is divided into multiple regions with distinct dynamic models, causing traditional data-driven methods to produce inadequate over-approximations of reachable sets at region boundaries where dynamics change abruptly. This paper introduces an approach using hybrid zonotopes for data-driven reachability analysis of piecewise affine systems. Our method addresses the boundary transition problem by developing computational algorithms that calculate the family of set models guaranteed to contain the true system trajectories. Additionally, we extend and evaluate three methods for set-based estimation that account for input-output data with measurement noise.
\end{abstract}

\section{Introduction}

Reachability analysis addresses a fundamental question in hybrid systems: can system states enter unsafe regions under given initial conditions and uncertainties? This question is central to ensuring safety in critical applications such as autonomous vehicles, medical devices, and industrial control systems. Hybrid systems, which combine continuous dynamics with discrete mode switching, present unique verification challenges that are not found in purely continuous or discrete systems. At region boundaries where dynamics change abruptly, traditional reachability methods often produce inaccurate approximations, leading to either false alarms or missed safety violations.

Previous approaches to hybrid system verification include Asarin et al.'s \cite{asarin2000approximate} algorithms for piecewise-linear systems and the zonotope/hyperplane intersection method by Girard et al. \cite{guernic2010zonotope}. While these methods established important foundations, they rely on known mathematical models, which are often unavailable in practice. Hybrid systems can be mathematically formulated in several equivalent ways, including Piecewise Affine (PWA) systems and mixed logical dynamical models \cite{bemporad2000observability}, with Heemels et al. \cite{heemels2001equivalence} establishing their theoretical unification.

Although exact computation of reachable sets for hybrid systems is generally undecidable, approximation techniques using zonotopes, introduced by Girard et al. \cite{girard2005reachability}, have proven effective for linear systems with uncertainty. More recently, Alanwar et al. \cite{amr23reachable} demonstrated data-driven reachability analysis, computing reachable sets directly from noisy measurements without prior model knowledge. However, their framework, while successful for linear and general nonlinear systems, does not address the critical boundary transition problem that is unique to hybrid systems.

PWA system identification faces three major interconnected challenges: determining the number of submodels, estimating affine parameters, and identifying partitioning hyperplanes \cite{paoletti2007identification}. Our research specifically targets the gap in reachability analysis at mode transitions by introducing hybrid zonotopes that can effectively capture the complex behavior of PWA systems across region boundaries.

This paper makes the following contributions:
\begin{itemize}
    \item  A data-driven method using hybrid zonotopes to compute reachable sets of PWA systems directly from noisy measurements
\item An algorithm that explicitly handles multiple operating modes and state-space partitioning, maintaining safety guarantees across transitions
\item A comprehensive set-based state estimation framework for Multiple-Input Multiple-Output (MIMO) Piecewise Affine systems with measurement noise
\item Three mathematically equivalent techniques for processing measurement noise, with a formal proof of their equivalence under clearly defined conditions
\end{itemize}

Our code is available online to recreate our findings\footnote{\href{https://github.com/TUM-CPS-HN/Data-Driven-Reachability-Analysis-for-Piecewise-Affine-System}{https://github.com/TUM-CPS-HN/Data-Driven-Reachability-Analysis-for-Piecewise-Affine-System}}. The remainder of this paper is organized as follows: Section \ref{sec:preliminaries} introduces hybrid zonotope representations and formally defines the problem statement. In Section \ref{state-input}, we present our approach for computing reachable sets of PWA systems from input-state data with process noise. Section \ref{setbased} extends our framework to handle input-output data with both process and measurement noise through online set-based estimation. Section \ref{sec:evaluation} demonstrates our methods through numerical examples. Finally, Section \ref{sec:conclusion} summarizes our contributions and discusses future research directions.

\section{Preliminaries and Problem Statement}\label{sec:preliminaries}

\textbf{Notation}: Throughout this paper, matrices are denoted by capital letters (e.g., $A$, $B$), identity matrix is represented by $I$, vectors by lowercase letters (e.g., $x$, $c$), and sets by calligraphic letters (e.g., $\mathcal{Z}$, $\mathcal{C}$). The set of real numbers is denoted by $\mathbb{R}$, with $\mathbb{R}^n$ representing $n$-dimensional Euclidean space. The pseudoinverse of an interval matrix is computed by following \cite{fiedler2006linear} and is denoted by $\dagger$. Time indices appear in parentheses (e.g., $x(k)$). Additional notation will be introduced as needed throughout the paper.

We begin by introducing the hybrid zonotope representation.
\subsection{Hybrid Zonotope Representations}
\begin{definition}[Hybrid Zonotope \cite{bird2023hybrid}]
A set $\mathcal{Z}_{h} \subset \mathbb{R}^{n}$ is a hybrid zonotope if there exist matrices $G^{c} \in \mathbb{R}^{n \times n_{g}}$, $G^{b} \in \mathbb{R}^{n \times n_{b}}$, $A^{c} \in \mathbb{R}^{n_{c} \times n_{g}}$, $A^{b} \in \mathbb{R}^{n_{c} \times n_{b}}$, and vectors $c \in \mathbb{R}^{n}$, $b \in \mathbb{R}^{n_{c}}$ such that
\begin{small}
\begin{equation}
\mathcal{Z}_h=\!\left\{\!\begin{array}{l}\!\left[G_{z}^c\;G_{z}^b\right]\!\left[\!\begin{array}{c}\!\xi^c\!\\\!\xi^b\!\end{array}\!\right]\!+\!c_z\!\left\lvert\,\!\begin{array}{c}\!{\left[\!\begin{array}{c}\!\xi^c\!\\\!\xi^b\!\end{array}\!\right]\!\in\!\mathcal{B}_{\infty}^{n_g}\!\times\!\{-1,1\}^{n_b},}\!\\{\left[A_{z}^c\;A_{z}^b\right]\!\left[\!\begin{array}{l}\!\xi^c\!\\\!\xi^b\!\end{array}\!\right]\!=\!b_z}\!\end{array}\!\right.\!\end{array}\!\right\}.
\end{equation}
\end{small}
\end{definition}
This is denoted concisely as hybrid zonotope, $\mathcal{Z}_{h}=\left\langle G_{z}^{c}, G_{z}^{b}, c_{z}, A_{z}^{c}, A_{z}^{b}, b_{z}\right\rangle$
\medskip

\begin{compactproposition}(Hybrid Zonotope Operations \cite{bird2023hybrid})
For hybrid zonotopes $Z_1=\left\langle G_{1}^{c}, G_{1}^{b}, c_{1}, A_{1}^{c}, A_{1}^{b}, b_{1}\right\rangle \subset \mathbb{R}^{n}$, $Z_2=\left\langle G_{2}^{c}, G_{2}^{b}, c_{2}, A_{2}^{c}, A_{2}^{b}, b_{2}\right\rangle \subset \mathbb{R}^{n}$, $Z_3=\left\langle G_{3}^{c}, G_{3}^{b}, c_{3}, A_{3}^{c}, A_{3}^{b}, b_{3}\right\rangle \subset \mathbb{R}^{m}$, a linear map $R \in \mathbb{R}^{m \times n}$, and a halfspace $\mathcal{H}^{-}=\left\{x \in \mathbb{R}^{m} \mid l^{T} x \leq \rho\right\}$, the following operations are defined:
\begin{small}
\begin{equation}
\begin{split}
Z_1 \oplus Z_2=\big\langle & [G_{1}^{c} \; G_{2}^{c}], [G_{1}^{b} \; G_{2}^{b}], c_{1}+c_{2}, \\
& \begin{bmatrix} A_{1}^{c} & \mathbf{0} \\ \mathbf{0} & A_{2}^{c} \end{bmatrix}, 
\begin{bmatrix} A_{1}^{b} & \mathbf{0} \\ \mathbf{0} & A_{2}^{b} \end{bmatrix}, 
\begin{bmatrix} b_{1} \\ b_{2} \end{bmatrix}\big\rangle,
\end{split}
\end{equation}
\begin{equation}
\begin{split}
Z_1 \cap_{R} Z_3=\big\langle & [G_{1}^{c} \; \mathbf{0}], [G_{1}^{b} \; \mathbf{0}], c_{1}, \begin{bmatrix} A_{1}^{c} & \mathbf{0} \\ \mathbf{0} & A_{3}^{c} \\ R G_{1}^{c} & -G_{3}^{c} \end{bmatrix},\\
&\begin{bmatrix} A_{1}^{b} & \mathbf{0} \\ \mathbf{0} & A_{3}^{b} \\ R G_{1}^{b} & -G_{3}^{b} \end{bmatrix},
\begin{bmatrix} b_{1} \\ b_{3} \\ c_{3}-R c_{1} \end{bmatrix}\big\rangle,
\end{split}
\label{hybridintersectionhybrid}
\end{equation}
\begin{equation}\label{halfspacehybridzono}
\begin{split}
Z_1 \cap_{R} \mathcal{H}^{-}=\big\langle & [G_{1}^{c} \; \mathbf{0}], G_{1}^{b}, c_{1}, \begin{bmatrix} A_{1}^{c} & \mathbf{0} \\ l^{T} R G_{1}^{c} & \frac{d_{m}}{2} \end{bmatrix},\\
&\begin{bmatrix} A_{1}^{b} \\ l^{T} R G_{1}^{b} \end{bmatrix},
\begin{bmatrix} b_{1} \\ \rho-l^{T} R c_{1}-\frac{d_{m}}{2} \end{bmatrix}\big\rangle,
\end{split}
\end{equation}
with $d_m=\rho-l^TR c_1+\sum_{i=1}^{n_{g,1}}|l^TRg_1^{(c,i)}|+\sum_{i=1}^{n_{b,1}}|l^TRg_1^{(b,i)}|$. 
\end{small}
\end{compactproposition}

\subsection{Problem Formulation}

We consider a discrete-time PWA system
\begin{subequations}\label{probformula}
\begin{align}
x(k+1) &= \begin{cases}
A_1x(k) + B_1u(k) + w(k) & \text{if } \delta_1(k)=1, \\
\,\,\,\,\,\, \,\,\, \,\,\, \,\,\, \,\,\, \,\,\, \,\,\, \,\,\, \,\,\,  \vdots & \\
A_sx(k) + B_su(k) + w(k) & \text{if } \delta_s(k)=1,
\end{cases} \label{pwastate} \\
y^{j}(k) &= C^{j} x(j)+v^{j}(k), \quad j \in\{1, \ldots, q\}. \label{pwaoutput}
\end{align}
\end{subequations}
Here, $x(k) \in \mathbb{R}^{n}$ represents the system state, $u(k) \in \mathbb{R}^{m}$ the control input, and $y^{j}(k) \in \mathbb{R}^{p_{j}}$ the measurement from sensor $j$. The system matrices $A_i \in \mathbb{R}^{n \times n}$ and $B_i \in \mathbb{R}^{n \times m}$ are unknown, while the observation matrices $C^{j} \in \mathbb{R}^{p_{j} \times n}$ are known for all sensors.
The binary variables $\delta_i(k) \in \{0,1\}$ indicate the active operating mode and satisfy the exclusive-or condition:
\begin{equation}
\Sigma_{i=1}^s\delta_i(k)=1.
\end{equation}
The process noise is defined as $w(k) \in \mathcal{Z}_{w}=\left\langle c_{w}, G_{w}\right\rangle \subset \mathbb{R}^{n}$. The measurement noise for each sensor $j$ is defined as $v^{j}(k) \in \mathcal{Z}_{v, j}=\left\langle c_{v, j}, G_{v, j}\right\rangle \subset \mathbb{R}^{p_{j}}$.

 For any $x$ from the state set and corresponding $y^j$, we can find a $\xi^j\in [-1,1]^{n_v}$ such that :
\begin{small}
\begin{equation}\label{eq:meas_noise_xi}
y^j = C^jx + v^j = C^jx + c_{v,j} + G_{v,j}\xi^j.
\end{equation}
\end{small}

For the system, the state space $\mathcal{C}$ must be partitioned into $s$ disjoint regions, and this partition is known, satisfying:
\begin{equation}
\mathcal{C}_i \cap \mathcal{C}_k = \emptyset, \quad \forall i \neq k,\quad \text{and}\quad \bigcup_{i=1}^s \mathcal{C}_i = \mathcal{C}.
\end{equation}
Each region $\mathcal{C}_i$ is defined as a polyhedral set given by linear inequalities:
\begin{equation}\label{region}
\mathcal{C}_i = \{\, x \in \mathbb{R}^n \mid L_i x \leq \rho_i \,\}.
\end{equation}
where $L_i \in \mathbb{R}^{m_i \times n}$ and $\rho_i \in \mathbb{R}^{m_i}$. Here, $L_i^{(j)}$ denotes the $j$-th row of $L_i$. 

The problem is how to over-approximate the model sets by using the above data in hybrid systems.

\section{Reachability of PWA System from Input-State Data}\label{state-input}
In this section, we address computing reachable sets of PWA systems when input-state data contains process noise. We establish methods to bound possible system states under uncertainty, providing the foundation for the more complex scenario of input-output data with both process and measurement noise, which we explore in the following section.
\subsection{Family of Set Models}
To over-approximate the PWA system’s behavior, we first partition the available data into \(s\) subsets according to \eqref{pwastate}. Then, for each operating mode \(i\), we consider input-state trajectories of length \(T_i + 1\). The data matrices for the $i$-th mode are constructed as:
\begin{equation}
X_i = [x_i(0) \ldots x_i(T_i)],
\end{equation}
The corresponding shifted signals for all modes are defined as:
\begin{equation}
\begin{aligned}
X_{i,+} &= [x_i(1) \ldots x_i(T_i)], \\
X_{i,-} &= [x_i(0) \ldots x_i(T_i-1)], \\
U_{i,-} &= [u_i(0) \ldots u_i(T_i-1)],\\
Y_{i,+} &= [y_i(1) \,\ldots\, y_i(T_i)], \\
Y_{i,-} &= [y_i(0) \,\ldots\, y_i(T_i-1)].
\label{inputandstate}
\end{aligned}
\end{equation}

The process noise sequence is denoted as:
\begin{equation}
W^{i,-}= \left[w(0) \ldots w(T_i-1)\right] \in \mathcal{M}_{w,i}.
\end{equation}
where $T = \sum_{i=1}^s T_i$ is the total number of time steps across all modes, and $\mathcal{M}_{w,i}$ is a matrix zonotope defined as:
\begin{equation}
\mathcal{M}_{w, i}=\left\langle C_{\mathcal{M}, w,i}, \tilde{G}_{\mathcal{M},i}\right\rangle.
\end{equation}
We extend \cite[Theorem 1]{alanwar2022data} to hybrid systems.
\begin{theorem}For a PWA system with $s$ modes, given the matrix zonotopes  $\mathcal{M}_{w,i}$ and input-state trajectories $D_i=(U_{i,-}, X_i)$ such that $\left[\begin{array}{ll}X_{i,-}^{\top} & U_{i,-}^{\top}\end{array}\right]^{\top}$ has full row rank, the matrix zonotope set is defined as:\label{theorem:familyset}
\begin{equation}
\mathcal{M}_{\Sigma} = \{\mathcal{M}_{\Sigma,i}\}_{i=1}^s = \Bigl\{\bigl\langle C_{\mathcal{M}_{\Sigma},i}, \tilde{G}_{\mathcal{M}_{\Sigma},i}\bigr\rangle\Bigr\}_{i=1}^s,
\end{equation}
where
\begin{equation}
\mathcal{M}_{\Sigma,i}=(X_{i,+}-\mathcal{M}_{w,i})\begin{bmatrix}
X_{i,-} \\
U_{i,-}
\end{bmatrix}^{\dagger},
\end{equation}
contains the complete system truth model set $\{[A_i \quad B_i]\}_{i=1}^s \subseteq \mathcal{M}_{\Sigma}$.
\end{theorem}

\begin{compactproof}
The proof is a straight forward extension of~\cite[Lemma 1]{amr23reachable}.
\end{compactproof}

For PWA systems with process noise, we compute the reachable state set by considering all possible mode transitions through three sequential operations:
\begin{subequations}
\begin{align}
\tilde{\mathcal{R}}_{k,i} &= \tilde{\mathcal{R}}_{k} \cap \mathcal{C}_i, \quad i = 1,\dots,s , \label{eq:a}\\[4pt]
\tilde{\mathcal{R}}_{k+1} &= \bigcup\nolimits_{i=1}^s \left(\mathcal{M}_{\Sigma,i} \left(\tilde{\mathcal{R}}_{k,i} \times \mathcal{U}_{k,i}\right) + \mathcal{Z}_w\right), \label{eq:b}\\[4pt]
\tilde{\mathcal{R}}_{k+1,i} &= \tilde{\mathcal{R}}_{k+1} \cap \mathcal{C}_i, \quad i = 1,\dots,s. \label{eq:c}
\end{align}
\label{pwaeq}
\end{subequations}
In these equations, $\tilde{\mathcal{R}}_{k,i}$ represents the state set in region $i$ after the previous update, $\mathcal{U}_{k,i}$ denotes the input set for submodel $i$, and $\mathcal{Z}_w$ is the process noise zonotope. \\

\begin{algorithm}[htb]
\caption{PWA Systems Reachability}\label{alg:pwa-reachability}
\begin{algorithmic}[1]
\REQUIRE input-state trajectories $D=\left(U_{i, -}, X_{i}\right)$, initial set $\mathcal{X}_0$, process noise zonotope $\mathcal{Z}_w$, matrix zonotope $\mathcal{M}_{w,i}$, input zonotope $\mathcal{U}_k, \forall k=0, \ldots, N-1$, state-space partitions $\mathcal{C}_i$ with $L_i, \rho_i $
\ENSURE reachable sets $\tilde{\mathcal{R}}_k, \forall k=1, \ldots, N$
\FOR{$i = 1$ \TO $s$}
    \STATE $\mathcal{M}_{\Sigma,i} = (X_{i,+} - \mathcal{M}_{w,i})\begin{bmatrix} X_{i,-} \\ U_{i,-} \end{bmatrix}^{\dagger}$
\ENDFOR
\STATE $\mathcal{M}_{\Sigma}  = \{\,\mathcal{M}_{\Sigma,i} \}_{i=1}^s$
\STATE $\tilde{\mathcal{R}}_0=\mathcal{X}_0$
\FOR{$k = 0$ \TO $N-1$}
    \FOR{$i = 1$ \TO $s$}
        \STATE $\tilde{\mathcal{R}}_{k,i}^{(0)} = \tilde{\mathcal{R}}_{k}$
        \FOR{$j = 1$ \TO $m_i$}
            \STATE $\tilde{\mathcal{R}}_{k,i}^{(j)} = \tilde{\mathcal{R}}_{k,i}^{(j-1)} \cap \{x \mid (L_i^{(j)})^T x \leq \rho_i^{(j)}\}$
        \ENDFOR
        \STATE $\tilde{\mathcal{R}}_{k,i} = \tilde{\mathcal{R}}_{k,i}^{(m_i)}$
    \ENDFOR
    \STATE $\tilde{\mathcal{R}}_{k+1} =  \bigcup\nolimits_{i=1}^s \biggl( \mathcal{M}_{\Sigma,i} \bigl(\tilde{\mathcal{R}}_{k,i} \times U_{k,i}\bigr) + \mathcal{Z}_w \biggr)$
\ENDFOR
\RETURN $\tilde{\mathcal{R}}_k$
\end{algorithmic}
\end{algorithm}

\begin{figure}[htbp]
\centering
\includegraphics[width=0.3\textwidth]{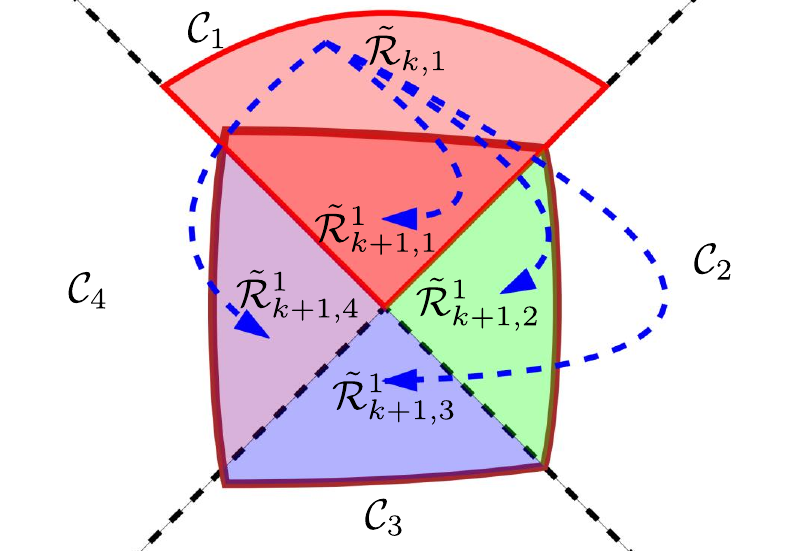}
\caption{State space partition and reachable set evolution of a PWA system. The figure illustrates the state space divided into four regions ($\mathcal{C}_1$, $\mathcal{C}_2$, $\mathcal{C}_3$, and $\mathcal{C}_4$) by diagonal dashed lines. The red curve encloses  $\tilde{\mathcal{R}}_{k,1}$ which is the reachable set at time $k$, intersecting with region $\mathcal{C}_1$. The solid dark red boundary encompasses the one-step reachable set $\tilde{\mathcal{R}}_{k+1}^1$ computed from $\tilde{\mathcal{R}}_{k,1}$. 
}
\label{fig:pwa_region}
\end{figure}

Fig.~\ref{fig:pwa_region} illustrates this process for a system with four regions. The propagation from region $\mathcal{C}_1$ might intersect with other regions, necessitating a comprehensive approach. Starting with a reachable set $\tilde{\mathcal{R}}_{k,1}$ in region $\mathcal{C}_1$, we compute the one-step reachable set $\tilde{\mathcal{R}}_{k+1}^1$ of $\tilde{\mathcal{R}}_{k,1}$ using ~\eqref{eq:b}, where the superscript $1$ indicates that these sets evolved from the previous set in region $\mathcal{C}_1$. This set crosses multiple region boundaries, resulting in subsets $\tilde{\mathcal{R}}_{k+1,i}^1$ for $i=1,\dots,4$ after applying~\eqref{eq:c}. 

Algorithm~\ref{alg:pwa-reachability} outlines the process for calculating the reachable set. We begin by representing the state set 
\(\tilde{\mathcal{R}}_{k}\) as a hybrid zonotope:
\[
\tilde{\mathcal{R}}_{k} 
= \left\langle G_z^c, G_z^b, c_z, A_z^c, A_z^b, b_z \right\rangle 
\subset \mathbb{R}^n.
\]
With the definition of each regions in \eqref{region}, we obtain the intersection result using~\eqref{halfspacehybridzono}.
By extending the  ~\cite[proposition 2]{amr23reachable} to handle 
hybrid zonotopes and matrix zonotope products, 
we derive the expression for 
\(\mathcal{M}_{\Sigma,i}\bigl(\tilde{\mathcal{R}}_{k,i} \times \mathcal{U}_{k,i}\bigr)\)
in each submodel~\(i\). 
Taking the union of these expressions across all submodels 
then yields the complete reachable set 
\(\tilde{\mathcal{R}}_{k+1}\), as specified in~\eqref{eq:b}.

\section{Online Set-Based Estimation}\label{setbased}
An approach for set-based estimation was presented in \cite{alanwar2022data} to handle input-output data corrupted by measurement noise. This approach involves two primary steps—time update and measurement update—to compute the reachable set for linear systems. In this work, we extend that methodology to MIMO PWA systems, addressing a key limitation of the original method, which fails under hybrid dynamics. 

\subsection{Time Update}
Given the measurement trajectories $Y_{i,+}$ and $Y_{i,-}$ along with the noise zonotope $\mathcal{Z}_{v,j}$ from \eqref{inputandstate}, we can compute the family set of models for the PWA system with measurement noise matrix zonotope, denoted by $\mathcal{M}_{\Sigma,i}^y$, by applying Theorem~\ref{theorem:familyset} and \cite[Lemma~1]{alanwar2022data}. Subsequently, implementing Algorithm~\ref{alg:pwa-reachability} yields the reachable set $\tilde{\mathcal{R}}_{k,i}^y$ that properly accounts for measurement noise.

\subsection{Measurement Update}
For PWA systems with multiple operating modes and noisy sensor measurements, 
we adapt three measurement-update methods to the hybrid setting: 
Reverse-Mapping (RM) from~\cite{alanwar2022data}, 
Implicit Intersection (IN) from~\cite{alanwar2022data}, 
and Generalized Intersection (GI) from~\cite{scott2016constrained}. 
These methods incorporate measurement information into the estimation process 
to compute sets of possible system states. 

For each sensor \(j \in \{1,\dots,q\}\), we employ the measurement equation 
\eqref{pwaoutput}. The final set-based state estimate is then obtained by intersecting 
the time-updated set with all measurement-based sets:
\begin{small}
\begin{equation}
    \hat{\mathcal{R}}_{k,i}^y 
    = \tilde{\mathcal{R}}_{k,i}^y 
    \,\cap\, 
    \biggl(\,\bigcap_{j=1}^q \mathcal{Z}_{x\mid y^j}(k)\biggr),
\end{equation}
\end{small}\noindent 
where \(\mathcal{Z}_{x\mid y^j}(k)\) is the state zonotope corresponding 
to the measurement data from sensor~\(j\):
\begin{small}
\begin{equation}
\begin{aligned}
    \mathcal{Z}_{x \mid y^j}(k)
    &= \bigl\{\,x \in \mathbb{R}^n 
    \mid C^j x = y^j(k) - \mathcal{Z}_{v,j}\bigr\} \\[5pt]
    &= \bigl\langle c_{x\mid y^j},\, G_{x\mid y^j}\bigr\rangle.
\end{aligned}
\end{equation}
\end{small}\noindent 
We obtain the predicted state set \(\tilde{\mathcal{R}}_{k,i}^y\) 
from the time-update step, representing it as a hybrid zonotope for each mode \(i\). We have 
\begin{equation}
    \tilde{\mathcal{R}}_{k,i}^y  =  \langle G_{k,i}^c,\, G_{k,i}^b,\, c_{k,i},\, A_{k,i}^c,\, A_{k,i}^b,\, b_{k,i}\rangle \subset \mathbb{R}^n.\nonumber
\end{equation}
Given the predicted state set \(\tilde{\mathcal{R}}_{k,i}^y\) and bounded-noise measurements, the corrected state-estimation set can be computed using any of the three approaches summarized in Algorithm~\ref{alg:threemehods}.

The RM approach addresses the state estimation problem by explicitly computing the state-space representation of measurement constraints. This method uses singular value decomposition (SVD) to construct measurement-compatible state zonotopes and then intersects them with the predicted state set.
\begin{compactproposition}[Reverse-Mapping]\label{RM} 
Given the predicted state set 
$\tilde{\mathcal{R}}_{k,i}^y = \langle G_{k,i}^c,\, G_{k,i}^b,\, c_{k,i},\, A_{k,i}^c,\, A_{k,i}^b,\, b_{k,i}\rangle$ 
and the measurement-based zonotope $\mathcal{Z}_{x\mid y^j}(k) = \langle c_{x\mid y^j},\, G_{x\mid y^j}\rangle$ from sensor $j$, 
the corrected state-estimation set at time step \(k\) for mode \(i\) is computed as follows:
\begin{footnotesize}
\begin{equation}\label{eq:rm_method}
\textstyle
\begin{aligned}
&\hat{\mathcal{R}}_{k+1,i} = \tilde{\mathcal{R}}_{k,i}^y \cap \Biggl( \bigcap\nolimits_{j=1}^{q} \mathcal{Z}_{x|y^j}(k) \Biggr)  \\[2mm]
& = \Big\langle 
\begin{bmatrix} G_{k,i}^c & \mathbf{0} & \cdots & \mathbf{0} \end{bmatrix}, 
\begin{bmatrix} G_{k,i}^b & \mathbf{0} & \cdots & \mathbf{0} \end{bmatrix}, 
c_{k,i}, \\
&\begin{bmatrix} 
A_{k,i}^c & \mathbf{0} & \cdots & \mathbf{0} \\
\vdots & \vdots & \ddots & \vdots \\
G_{k,i}^c & -G_{x|y^1} & \cdots & \mathbf{0} \\
\vdots & \vdots & \ddots & \vdots \\
G_{k,i}^c & \mathbf{0} & \cdots & -G_{x|y^q}
\end{bmatrix},\begin{bmatrix}
A_{k,i}^b & \mathbf{0} & \cdots & \mathbf{0} \\
\vdots & \vdots & \ddots & \vdots \\
G_{k,i}^b & \mathbf{0} & \cdots & \mathbf{0} \\
\vdots & \vdots & \ddots & \vdots \\
G_{k,i}^b & \mathbf{0} & \cdots & \mathbf{0}
\end{bmatrix}, 
\begin{bmatrix}
b_{k,i} \\
\vdots \\
c_{x|y^1} - c_{k,i} \\
\vdots \\
c_{x|y^q} - c_{k,i}
\end{bmatrix}
\Big\rangle,
\end{aligned}
\end{equation}
\end{footnotesize}
where for each $j \in \{1,\ldots,q\}$ :
\begin{small}
\begin{align}
c_{x|y^j} &= V^j_1 \Sigma^{-1} P_1^\top\left(y^j(k)-c_{v,j}\right) ,\label{eq:rm_center}\\
G_{x|y^j} &= \left[V^j_1 \Sigma^{-1} P_1^\top G_{v,j} \quad V^j_2 M\right] .\label{eq:rm_generator}
\end{align}  
\end{small}\noindent 
with $P^j_1$, $\Sigma$, $V^j_1$, and $V^j_2$ obtained from the SVD of $C^j$. The parameter $M$ is chosen sufficiently large to account for uncertainties in the null space.
\end{compactproposition}

\begin{compactproof}
For each measurement $y^j(k)$, we construct a state-space zonotope $\mathcal{Z}_{x|y^j}(k) = \left\langle c_{x|y^j}, G_{x|y^j}\right\rangle$ through the SVD of $C^j = P_1 \Sigma V_1^\top$. Setting $\mathcal{Z}_1 \equiv \tilde{\mathcal{R}}_{k,i}^y$ and $\mathcal{Z}_3 \equiv \mathcal{Z}_{x|y^j}(k)$ to \eqref{hybridintersectionhybrid} to obtain:
\begin{small}
\begin{equation}\label{eq:svd_single}
\begin{aligned}
\hat{\mathcal{R}}_{k,i}^{y,\text{RM}} &= \tilde{\mathcal{R}}_{k,i}^y \cap\mathcal{Z}_{x|y^j}(k) \\
&= \big\langle  [{G}_{k,i}^c \; \mathbf{0}], [{G}_{k,i}^b \; \mathbf{0}], {c}_{k,i}, \begin{bmatrix} {A}_{k,i}^c & \mathbf{0} \\ \mathbf{0} & \mathbf{0} \\ {G}_{k,i}^c &  -G_{x|y^j} \end{bmatrix},\\
&\begin{bmatrix} {A}_{k,i}^b & \mathbf{0} \\ \mathbf{0} & \mathbf{0} \\ {G}_{k,i}^b & \mathbf{0} \end{bmatrix},
\begin{bmatrix} {b}_{k,i} \\ \mathbf{0} \\ c_{x|y^j} - c_{k,i} \end{bmatrix}\big\rangle.
\end{aligned}
\end{equation}
\end{small}\noindent 
Extending to multiple measurements by sequential application of the intersection formula yields the final result in \eqref{eq:rm_method}. 
\end{compactproof}

Contrary to the RM approach, in the IN method, we do not explicitly determine the sets $\mathcal{Z}_{x \mid y^i(k)}$. Instead, $\hat{\mathcal{R}}_{k,i}^{y,\text{IN}}$ is determined directly from the set $\tilde{\mathcal{R}}^y_{k,i}$, the measurements $y^i(k)$ and some weights $\lambda_k^i$ for $i \in$ $\{1, \ldots, q\}$. We then optimize over the weights to minimize the volume of $\hat{\mathcal{R}}_{k,i}^{y,\text{IN}}$, resulting in a more compact representation of the state-estimate set.
\begin{compactproposition}[Implicit Intersection]\label{IN}
Given the predicted state set $\tilde{\mathcal{R}}_{k,i}^y = \langle G_{k,i}^c,\, G_{k,i}^b,\, c_{k,i},\, A_{k,i}^c,\, A_{k,i}^b,\, b_{k,i}\rangle$  for mode $i$ and the $q$ regions for $x$ corresponding to $y^j(k)$ with noise $v^j(k) \in \mathcal{Z}_{v,j} = \langle{c_{v,j},G_{v,j}}\rangle$ satisfying \eqref{pwaoutput}, the corrected state-estimation set $\hat{\mathcal{R}}_{k,i}^{y,\text{IN}}=\left\langle \hat{G}_{k,i}^{c,\text{IN}}, \hat{G}_{k,i}^{b,\text{IN}}, \hat{c}_{k,i}^\text{IN}, \hat{A}_{k,i}^{c,\text{IN}}, \hat{A}_{k,i}^{b,\text{IN}}, \hat{b}_{k,i}^\text{IN} \right\rangle$ can be computed:
\begin{footnotesize}
\begin{equation}\label{eq:ii_method}
\begin{aligned}
\hat{G}_{k,i}^{c,\text{IN}} &= \left[(I-\sum_{j=1}^q \lambda_k^j C^j){G}_{k,i}^c \quad -\lambda_k^1G_{v,1} \quad \cdots \quad -\lambda_k^qG_{v,q}\right], \\
\hat{G}_{k,i}^{b,\text{IN}} &= (I-\sum_{j=1}^q \lambda_k^j C^j){G}_{k,i}^b, \\
\hat{c}_{k,i}^\text{IN} &= {c}_{k,i} + \sum_{j=1}^q \lambda_k^j(y^j - C^j{c}_{k,i} - c_{v,j}), \\
\hat{A}_{k,i}^{c,\text{IN}} &= \begin{bmatrix} {A}_{k,i}^c & \underbrace{0 \quad \cdots \quad 0}_{q} \end{bmatrix}, \quad \hat{A}_{k,i}^{b,\text{IN}} = {A}_{k,i}^b, \quad  \hat{b}_{k,i}^\text{IN} = {b}_{k,i}.
\end{aligned}
\end{equation}
\end{footnotesize}
where\quad $\lambda_k^j \in \mathbb{R}^{n \times p_j}$ for $j \in\{1,\ldots,q\}$ are weight matrices.
\end{compactproposition}
 Like the RM method, the IN method also first compute measurements for $j$th individual sensors before combining them for all sensors, so the following discussion will focus only on the individual case.
\begin{compactproof}
 Let $x \in \tilde{\mathcal{R}}_{k,i}^y 
    \,\cap\, 
    \biggl(\,\bigcap_{j=1}^q \mathcal{Z}_{x\mid y^j}(k)\biggr)$, then there are  $\xi^c \in [-1,1]^{n_c}$ and $\xi^b \in \{-1,1\}^{n_b}$ to satisfy
\begin{equation}\label{eq:IN1}
x = {c}_{k,i} + {G}_{k,i}^c\xi^c + {G}_{k,i}^b\xi^b.
\end{equation}
To incorporate the measurement information optimally, we introduce weight matrices $\lambda_k^j \in \mathbb{R}^{n \times p_j}$ for each sensor $j$ from paper \cite{alanwar2023distributed}. 
Using these weight matrices, we add and subtract the term $\sum_{j=1}^q(\lambda_k^j C^j {G}_{k,i}^c \xi^c)$ into \eqref{eq:IN1}:
\begin{small}
\begin{equation}\label{eq:ii_term_add}
\begin{aligned}
x &= {c}_{k,i} + \sum_{j=1}^q(\lambda_k^j C^j {G}_{k,i}^c \xi^c) + (I-\sum_{j=1}^q \lambda_k^j C^j) {G}_{k,i}^c \xi^c + {G}_{k,i}^b\xi^b,
\end{aligned}
\end{equation}
\end{small}\noindent 
Putting \eqref{eq:IN1} into \eqref{eq:meas_noise_xi},
\begin{small}
\begin{equation}\label{eq:meas_eq}
y^j = C^j({c}_{k,i} + {G}_{k,i}^c\xi^c + {G}_{k,i}^b\xi^b) + c_{v,j} + G_{v,j}\xi^j,
\end{equation}
\end{small}\noindent 
Rearranging \eqref{eq:meas_eq} for $C^j {G}_{k,i}^c \xi^c$:
\begin{small}
\begin{equation}\label{eq:solve_cg}
C^j {G}_{k,i}^c \xi^c = y^j - C^j{c}_{k,i} - c_{v,j} - G_{v,j}\xi^j - C^j{G}_{k,i}^b\xi^b,
\end{equation}
\end{small}\noindent 
Substituting  \eqref{eq:solve_cg} into  \eqref{eq:ii_term_add} and rearranging terms, we obtain the optimal intersection representation:
\begin{footnotesize}
\begin{equation}\label{eq:ii_rep}
\begin{aligned}
x &= \underbrace{{c}_{k,i} + \sum_{j=1}^q \lambda_k^j(y^j - C^j{c}_{k,i} - c_{v,j})}_{\hat{c}_{k,i}^\text{IN}} \\
&+\underbrace{\left[(I-\sum_{j=1}^q \lambda_k^j C^j){G}_{k,i}^c \quad -\lambda_k^1G_{v,1} \quad \cdots \quad -\lambda_k^qG_{v,q}\right]}_{\hat{G}_{k,i}^{c,\text{IN}}}
\underbrace{\begin{bmatrix}\xi^c\\\xi^1\\\vdots\\\xi^q\end{bmatrix}}_{\hat{\xi}^{c}_\text{IN}} \\
&+\underbrace{(I-\sum_{j=1}^q \lambda_k^j C^j){G}_{k,i}^b}_{\hat{G}_{k,i}^{b,\text{IN}}}\underbrace{\xi^b}_{\hat{\xi}^{b}_\text{IN}}.
\end{aligned}
\end{equation}
\end{footnotesize}
The constraints transform accordingly to maintain the hybrid zonotope structure:
\begin{small}
\begin{equation}\label{eq:ii_constraints}
\begin{bmatrix}{A}_{k,i}^c & \mathbf{0}\end{bmatrix}\begin{bmatrix}\xi^c\\\xi^1\\\vdots\\\xi^q\end{bmatrix} + {A}_{k,i}^b\xi^b = {b}_{k,i}.
\end{equation}
\end{small}\noindent
Thus, we obtain an optimal hybrid zonotope representation of the intersection as shown in  \eqref{eq:ii_method}, where the weight matrices are chosen to minimize the generator matrix size while preserving all measurement constraints.
\end{compactproof}

These weights matrices are determined by solving the optimization problem:
\begin{equation}
\bar{\lambda}_k^* = \arg\min_{\bar{\lambda}_k} (\|\hat{G}_{k,i}^{c,\text{IN}}\|_F^2 + \alpha\|\hat{G}_{k,i}^{b,\text{IN}}\|_F^2 ).\label{optimaleq}
\end{equation}
where $\bar{\lambda}_k = [\lambda_k^1 \;\; \cdots \;\; \lambda_k^q]$ concatenates all weight matrices, $\alpha$ represents a positive real-valued weighting coefficient, and $\|\cdot\|_F$ denotes the Frobenius norm. This optimization ensures minimal over-approximation in the resulting intersection.

Unlike this optimization-based IN approach, the GI method instead provides a direct algebraic formulation by expressing measurement equations as additional constraints in the hybrid zonotope representation, maintaining a clear connection to the original system dynamics.\\

\begin{compactproposition}[Generalized Intersection]\label{GI}
Given the predicted state set $\tilde{\mathcal{R}}_{k,i}^y = \langle G_{k,i}^c,\, G_{k,i}^b,\, c_{k,i},\, A_{k,i}^c,\, A_{k,i}^b,\, b_{k,i}\rangle$  for mode $i$ and the $q$ regions for $x$ corresponding to $y^j(k)$ with noise $v^j(k) \in \mathcal{Z}_{v,j} = \langle{c_{v,j},G_{v,j}}\rangle$ satisfying \eqref{pwaoutput}, the corrected state set can be computed as follows.
\begin{small}
\begin{equation}\label{eq:gi_method}
\begin{split}
\hat{\mathcal{R}}_{k,i}^{y,\text{GI}} = \big\langle & 
\begin{bmatrix}
{G}_{k,i}^c & \mathbf{0}
\end{bmatrix}, 
\begin{bmatrix}
{G}_{k,i}^b & \mathbf{0}
\end{bmatrix},
{c}_{k,i}, \begin{bmatrix}
{A}_{k,i}^c & \mathbf{0} \\
\mathbf{0} & \mathbf{0} \\
C^j_i{G}_{k,i}^c & G_{v,j}
\end{bmatrix}, \\
&\begin{bmatrix}
{A}_{k,i}^b & \mathbf{0} \\
\mathbf{0} & \mathbf{0} \\
C^j_i{G}_{k,i}^b & \mathbf{0}
\end{bmatrix}, \begin{bmatrix}
{b}_{k,i} \\
\mathbf{0} \\
y^j(k) - c_{v,j}\; - C^j_i{c}_{k,i}
\end{bmatrix}
\big\rangle.
\end{split}
\end{equation}
\end{small}
\end{compactproposition}

\begin{compactproof}
Consider the PWA system defined in the system equations. According to\cite{scott2016constrained}, the true state set $\hat{X}_k$ satisfies the recursive relation:
\begin{small}
\begin{equation}\label{eq:true_state_set}
\begin{gathered}
\hat{X}_k=\left(\left[A_i\; B_i\right] (\hat{X}_{k-1} \times \mathcal{U}_{k-1})+ \mathcal{Z}_w\right) \cap_{C^j_i}\left(y^j(k)-v^j\right), \\
\text{with} \quad \hat{X}_0=X_0 \cap_{C^j_i}\left(y^j(0)-v^j\right).
\end{gathered}
\end{equation}
   
\end{small}\noindent 
Since the exact computation of these set operations is generally intractable, with $\tilde{\mathcal{R}}_{k,i}^{y} \supseteq \hat{X}_k$, $\mathcal{M}^y_{\Sigma,i} \supseteq [A_i \quad B_i]$,
the set-based estimation can then be refined through measurement updates:
\begin{small}
\begin{equation}\label{eq:gi_update}
\begin{gathered}
\hat{\mathcal{R}}_{k,i}^{y,\text{GI}} = \left(\mathcal{M}^y_{\Sigma,i} \left(\tilde{\mathcal{R}}^y_{k,i} \times \mathcal{U}_{k,i}\right) + \mathcal{Z}_w\right) \cap_{C^j_i}\left(y^j(k)-v^j\right), \\
\text{with} \quad \hat{\mathcal{R}}_{0,i}^{y,\text{GI}}=\mathcal{R}_0 \cap_{C^j_i}\left(y^j(0)-v^j\right).
\end{gathered}
\end{equation}
\end{small}
By letting $\mathcal{Z}_1 = \tilde{\mathcal{R}}_{k,i}^y$, $\mathcal{Z}_3 = y^j(k) - v^j$, and $R = C^j_i$, and applying the hybrid zonotope intersection formula, we obtain the generalized intersection formulation shown in  \eqref{eq:gi_method}.
\end{compactproof}

\begin{algorithm}[htb]
\caption{Three Set-Based State Estimation Approaches}
\begin{algorithmic}\label{alg:threemehods}
\REQUIRE $\tilde{\mathcal{R}}_{k,i}^y, y^j(k), C^j, \mathcal{Z}_{v,j}$ for $j \in \{1,\ldots,q\}$
\ENSURE $\hat{\mathcal{R}}_{k,i}^y$
\IF{Approach 1 (Reverse-Mapping)}
    \FOR{$j = 1$ \TO $q$}
        \STATE Compute SVD: $C^j = P_1\Sigma V_1^\top$
        \STATE $c_{x|y^j} = V^t_1 \Sigma^{-1} P_1^\top(y^j(k)-c_{v,j})$
        \STATE $G_{x|y^j} = [V^t_1 \Sigma^{-1} P_1^\top G_{v,j} \quad V^t_2 M]$
    \ENDFOR
    \STATE Compute $\hat{\mathcal{R}}_{k,i}^y$ using~\eqref{eq:svd_single}
\ELSIF{Approach 2 (Implicit Intersection)}
    \FOR{$j = 1$ \TO $q$}
        \STATE Compute $\lambda_k^j$ using~\eqref{optimaleq}
    \ENDFOR
    \STATE Compute $\hat{\mathcal{R}}_{k,i}^y$ using~\eqref{eq:ii_method}
\ELSIF{Approach 3 (Generalized Intersection)}
    \STATE Compute $\hat{\mathcal{R}}_{k,i}^y$ using~\eqref{eq:gi_method}
\ENDIF
\RETURN $\hat{\mathcal{R}}_{k,i}^y$
\end{algorithmic}
\end{algorithm}
While the three approaches appear quite different in their formulations, they actually represent the same set of feasible states under the conditions based on \eqref{probformula}. The following theorem establishes their mathematical equivalence, and in subsequent sections, we will conduct benchmark testing of these three methods and compare their computational efficiency.
\begin{theorem}[Equivalence of the RM, IN, GI Methods]\label{thm:equivalence}
The three intersection approaches—Reverse Mapping (RM), Implicit Intersection (IN), and Generalized Intersection (GI)—yield identical results when measurement matrices have full rank, optimal weight matrices are found, sufficient parameterization is used for null space components, and all constraints are linear with bounded noise (represented as zonotopes).
\end{theorem}

\begin{compactproof}


We start by proving the equivalence of GI and RM Approaches. First, we prove $\hat{\mathcal{R}}_{k,i}^{y,\text{GI}} 
\subseteq \hat{\mathcal{R}}_{k,i}^{y,\text{RM}}$. For any point $x \in \hat{\mathcal{R}}_{k, i}^\text{GI}$, by proposition \ref{GI}, there exist $\xi_\text{GI}^v \in[-1,1]^{n_v}$ such that:
\begin{small}
\begin{equation}\label{eq:gi_cjx}
C_i^j x = y^j(k) - c_{v, j} - G_{v, j} \xi_\text{GI}^v.
\end{equation}
\end{small}\noindent 
Using the SVD decomposition 
\begin{equation}
    C^j_i = P_1^j \Sigma^j V_1^{j^T}\label{svdcomp},
\end{equation}
We can rewrite  \eqref{eq:gi_cjx}:
\begin{small}
\begin{equation}\label{eq:gi_svd1}
P_1^j \Sigma^j V_1^{j^T} x 
= y^j(k) - c_{v,j} - G_{v,j}\,\xi_\text{GI}^v.
\end{equation}
\end{small}\noindent 
Left-multiplying  \eqref{eq:gi_svd1} by $\Sigma^{j^{-1}}P_1^{j^T}$ (noting $P_1^{j^T} P_1^j = I$):
\begin{small}
\begin{equation}\label{eq:gi_svd3}
V_1^{j^T} x 
= \Sigma^{j^{-1}}P_1^{j^T}\bigl(y^j(k) - c_{v,j} - G_{v,j}\,\xi_\text{GI}^v\bigr).
\end{equation}
\end{small}\noindent 
We can decompose any state vector $x$ into components within the range and null spaces of $C^j$:
\begin{small}
\begin{equation}\label{eq:x_decomp}
x = V_1^j V_1^{j^T} x + V_2^j V_2^{j^T} x.
\end{equation}
\end{small}\noindent 
This decomposition stems from \eqref{svdcomp}, where $V_1^j$ span the range space of $(C^j)^T$, while $V_2^j$ spans the null space of $C^j$, with properties $V_1^{j^T} V_2^j = 0$.
\noindent Substituting the expression for $V_1^{j^T} x$ from  \eqref{eq:gi_svd3} into  \eqref{eq:x_decomp}:
\begin{small}
\begin{equation}\label{eq:gi_full_x}
x 
= V_1^j \Sigma^{j^{-1}}P_1^{j^T}\bigl(y^j(k) - c_{v,j} - G_{v,j}\,\xi_\text{GI}^v\bigr)
+ V_2^j V_2^{j^T} x.
\end{equation}
\end{small}\noindent 
Setting $-\xi_\text{GI}^v = \xi^v$ and $V_2^{j^T} x = M\xi^{null}$ (with a sufficiently large $M$) into\eqref{eq:gi_full_x}:
\begin{small}
\begin{equation}\label{eq:gi_rm_conversion}
\begin{split}
x &= V_1^j \Sigma^{j^{-1}}P_1^{j^T}\bigl(y^j(k) - c_{v,j}\bigr) \\
&\quad + V_1^j \Sigma^{j^{-1}}P_1^{j^T} G_{v,j}\,\xi^v 
  + V_2^j M\,\xi^{null},
\end{split}
\end{equation}
\end{small}\noindent 
This can be rewritten as:
\begin{small}
\begin{equation}\label{eq:gi_as_rm}
\begin{split}
x =& \underbrace{V_1^j (\Sigma^j)^{-1}(P_1^j)^T\bigl(y^j(k) - c_{v,j}\bigr)}_{c_{x|y^j}} \\
& + \underbrace{\bigl[V_1^j (\Sigma^j)^{-1}(P_1^j)^T G_{v,j} 
\quad V_2^j M\bigr]}_{G_{x|y^j}}
\begin{bmatrix} \xi^v \\ \xi^{null} \end{bmatrix}.
\end{split}
\end{equation}
\end{small}\noindent 
We can rewrite  \eqref{eq:gi_as_rm} as:
\begin{small}
\begin{equation}\label{eq:gi_as_rm_final}
x 
= c_{x|y^j} 
+ G_{x|y^j}\begin{bmatrix} \xi^v \\ \xi^{null} \end{bmatrix}.
\end{equation}
\end{small}\noindent 
This demonstrates that the point $x$ lies within the measurement-based zonotope $\mathcal{Z}_{x|y^j}(k)$ because:
\begin{itemize}
\item $\xi^v = -\xi_\text{GI}^v \in [-1,1]^{n_v}$ (since $\xi_\text{GI}^v \in [-1,1]^{n_v}$, its negation remains in the same interval)
\item $\xi^{null} = \frac{1}{M}(V_2^j)^T x \in [-1,1]^{n_{null}}$ (for sufficiently large $M$)
\end{itemize}Furthermore, since $x \in \tilde{\mathcal{R}}_{k,i}^y$ (as $x$ was taken from the GI representation), we have:
\begin{small}
\begin{equation}\label{eq:gi_subset_rm}
x 
\in \tilde{\mathcal{R}}_{k,i}^y 
  \cap \mathcal{Z}_{x|y^j}(k) 
= \hat{\mathcal{R}}_{k,i}^{y,\text{RM}}.
\end{equation}
\end{small}\noindent 
Thus, we have proven that any point in the GI representation can be expressed in the RM representation, $\hat{\mathcal{R}}_{k,i}^{y,\text{GI}} 
\subseteq \hat{\mathcal{R}}_{k,i}^{y,\text{RM}}$.

We next prove that $\hat{\mathcal{R}}_{k,i}^{y,\text{RM}} 
\subseteq \hat{\mathcal{R}}_{k,i}^{y,\text{GI}}$. Consider $x \in \hat{\mathcal{R}}_{k,i}^{y,\text{RM}}$
There exists $\xi_\text{RM}^v$ and $\xi_\text{RM}^{null}$ such that the point $x$ can be represented within the measurement-based zonotope as:
\begin{small}
\begin{equation}\label{eq:rm_zonotope_rep}
x 
= c_{x|y^j} 
  + G_{x|y^j}\begin{bmatrix} \xi_\text{RM}^v \\ \xi_\text{RM}^{null} \end{bmatrix},
\end{equation}
\end{small}\noindent 
Given \eqref{svdcomp}, \eqref{eq:rm_center} and \eqref{eq:rm_generator}:
\begin{small}
\begin{align}
C^j_i c_{x|y^j} 
&= P_1^j \Sigma^j (V_1^j)^T V_1^j (\Sigma^j)^{-1} (P_1^j)^T(y^j(k)-c_{v,j}) \nonumber\\
&= P_1^j (P_1^j)^T(y^j(k)-c_{v,j}) 
= y^j(k)-c_{v,j}, \label{eq:cjc_property}
\end{align}
\end{small}\noindent 
Similarly:
\begin{small}
\begin{equation}\label{eq:cjg_property}
C^j_i G_{x|y^j} = [\,G_{v,j} \quad \mathbf{0}\,].
\end{equation}
\end{small}\noindent 
Therefore, combining \eqref{eq:rm_zonotope_rep}, \eqref{eq:cjc_property}, and \eqref{eq:cjg_property}:
\begin{align}
C^j_i x 
&= C^j_i\Bigl(c_{x|y^j} 
+ G_{x|y^j}\begin{bmatrix} \xi_\text{RM}^v \\ \xi_\text{RM}^{null} \end{bmatrix}\Bigr) \nonumber\\
&= (y^j(k)-c_{v,j}) + G_{v,j}\,\xi_\text{RM}^v. \label{eq:cjx_xi_v}
\end{align}\noindent 
Setting $-\xi^v = \xi_\text{RM}^v \in [-1,1]^{n_v}$ in  \eqref{eq:cjx_xi_v}, we obtain
\begin{align}
C^j_i x = y^j(k)-c_{v,j} - G_{v,j}\,\xi^v \in \left( y^j(k) -v^j \right), \label{eq:gi-rm}\\
\label{eq:gi-rm2}
x \in I \cap_{C^j_i}\left(y^j(k)-v^j\right).
\end{align}Since $x \in \tilde{\mathcal{R}}_{k,i}^y$ and \eqref{eq:gi-rm2}, we have $x \in \tilde{\mathcal{R}}_{k,i}^y\cap_{C^j_i}\left(y^j(k)-v^j\right) = \hat{\mathcal{R}}_{k,i}^{y,\text{GI}}$.
Therefore, $\hat{\mathcal{R}}_{k,i}^{y,\text{RM}} 
\subseteq \hat{\mathcal{R}}_{k,i}^{y,\text{GI}}.$
Combined with the previous result 
\(\hat{\mathcal{R}}_{k,i}^{y,\text{GI}} 
\subseteq \hat{\mathcal{R}}_{k,i}^{y,\text{RM}}\), we conclude 
\(\hat{\mathcal{R}}_{k,i}^{y,\text{GI}} 
= \hat{\mathcal{R}}_{k,i}^{y,\text{RM}}\).

Next, we examine the relationship between the IN method and the RM method.
We first prove $\hat{\mathcal{R}}_{k,i}^{y,\text{IN}} \subseteq \hat{\mathcal{R}}_{k,i}^{y,\text{RM}}$. Considering any point $x$ in the IN representation, from \eqref{optimaleq} we have:
\begin{small}
\begin{equation}
\bar{\lambda}_k^* = \arg\min_{\bar{\lambda}_k} (\|\hat{G}_{k,i}^{c,\text{IN}}\|_F^2 + \alpha\|\hat{G}_{k,i}^{b,\text{IN}}\|_F^2)
\label{eq:opt_problem}.
\end{equation}
\end{small}\noindent 
From the proposition \ref{IN}, we can find a $\xi_{\text{IN}}^v \in [-1,1]^{n_v}$ to satisfy:
\begin{small}
\begin{align}
 C^j x &= y^j - c_{v,j} - G_{v,j}\xi_{\text{IN}}^v \label{eq:meas_constraint}.
\end{align}
\end{small}\noindent 
To simplify notation, let $L_k = \sum_{j=1}^q \lambda_k^j C^j$.\\
With \eqref{eq:ii_method}, the objective function in \eqref{eq:opt_problem} can be rewritten as:
\begin{small}
\begin{equation}
J(\bar{\lambda}_k) = \|(I-L_k){G}_{k,i}^c\|_F^2 + \alpha\|(I-L_k){G}_{k,i}^b\|_F^2 + \sum_{j=1}^q \|\lambda_k^jG_{v,j}\|_F^2.
\label{eq:obj_function}
\end{equation}
\end{small}\noindent 
We compute the gradient of $J$ with respect to $\lambda_k^j$:
\begin{small}
\begin{align}
Y_j &= \frac{\partial J}{\partial \lambda_k^j} \nonumber = -2(C^j)^T[(I-L_k)G_{k,i}^c](G_{k,i}^c)^T\\
&- 2\alpha(C^j)^T[(I-L_k)G_{k,i}^b](G_{k,i}^b)^T + 2\lambda_k^jG_{v,j}G_{v,j}^T.
\label{eq:gradient}
\end{align}
\end{small}\noindent 
We apply \eqref{svdcomp} and substitute into $L_k$:
\begin{small}
\begin{equation}
L_k = \sum_{j=1}^q \lambda_k^j P_1^j \Sigma^j (V_1^j)^T
\label{eq:L_k_svd}
\end{equation}
\end{small}\noindent 
We consider a candidate solution:$\lambda_k^j = V_1^j (\Sigma^j)^{-1} (P_1^j)^T$, substitute into \eqref{eq:L_k_svd} using orthogonality $(P_1^j)^T P_1^j = I$:
\begin{small}
\begin{equation}
L_k = \sum_{j=1}^q V_1^j (V_1^j)^T =I.
\label{eq:L_k_simplified}
\end{equation}
\end{small}\noindent 
Substituting~\eqref{eq:L_k_simplified} into~\eqref{eq:gradient}, we find:
\begin{small}
\begin{equation}
Y_j = 2\lambda_k^jG_{v,j}G_{v,j}^T \neq 0.
\label{eq:nonzero_gradient}
\end{equation}
\end{small}\noindent 
Thus, while $\lambda_k^j = V_1^j (\Sigma^j)^{-1} (P_1^j)^T$ is not the gradient-optimal solution, the true optimal solution $\bar{\lambda}_k^*$ from \eqref{eq:opt_problem} necessarily provides a more compact bound. Here, we define a reachable set $\check{\mathcal{R}}_{\bar{\lambda}_k}$ dependent on the parament $\bar{\lambda}_k$.\\ 
Therefore, $$x \in \hat{\mathcal{R}}_{k,i}^{y,\text{IN}} = \check{\mathcal{R}}_{\bar{\lambda}_k^*}\subseteq \check{\mathcal{R}}_{V_1^j (\Sigma^j)^{-1} (P_1^j)^T}.$$Substituting $\lambda_k^j = V_1^j (\Sigma^j)^{-1} (P_1^j)^T$ into \eqref{eq:ii_rep}:
\begin{small}
\begin{align}
x = &\, c_{k,i} 
+ \sum_{j=1}^q 
  V_1^j\,(\Sigma^j)^{-1}\,(P_1^j)^T 
  \bigl(y^j - C^j c_{k,i} - c_{v,j}\bigr) \nonumber\\
&\quad + \Bigl(I - \sum_{j=1}^q 
  V_1^j\,(\Sigma^j)^{-1}\,(P_1^j)^T\,C^j\Bigr)G_{k,i}^c\,\xi_\text{IN}^c ,
\label{eq:state_expr}
\end{align}
\end{small}\noindent 
Substituting \eqref{svdcomp} into \eqref{eq:meas_constraint}:
\begin{small}
\begin{align}
P_1^j \Sigma^j (V_1^j)^T x &= y^j - c_{v,j} - G_{v,j}\xi_{\text{IN}}^v  \label{eq:svd_meas},
\end{align}
\end{small}\noindent 
Multiplying both sides by $(\Sigma^j)^{-1} (P_1^j)^T$:
\begin{small}
\begin{align}
(V_1^j)^T x &= (\Sigma^j)^{-1} (P_1^j)^T(y^j - c_{v,j} - G_{v,j}\xi_{\text{IN}}^v ) ,\label{eq:projected_x}
\end{align}
\end{small}\noindent 
Since $(V_1^j)^T$ only captures the part of $x$ in the range space of $(C^j)^T$, the general solution must include a component in the null space of $C^j$:
\begin{small}
\begin{align}
x &= V_1^j (\Sigma^j)^{-1} (P_1^j)^T(y^j - c_{v,j} - G_{v,j}\xi_{\text{IN}}^v ) + V_2^j\eta ,\label{eq:general_solution}
\end{align}
\end{small}\noindent 
where $V_2^j$ spans the null space of $C^j$ and $\eta$ is an arbitrary vector. Rearranging \eqref{eq:general_solution}:
\begin{small}
\begin{align}
x &= V_1^j (\Sigma^j)^{-1} (P_1^j)^T(y^j - c_{v,j}) - V_1^j (\Sigma^j)^{-1} (P_1^j)^T G_{v,j}\xi_{\text{IN}}^v  + V_2^j\eta \label{eq:rearranged}.
\end{align}
\end{small}\noindent 
Setting $\xi^v = -\xi_{\text{IN}}^v $ and $\xi^{null} = \eta/M$ where $M$ is sufficiently large, we arrive at the standard reverse mapping form:
\begin{small}
\begin{equation}\label{eq:gi_as_rm2}
\begin{split}
x =& \underbrace{V_1^j (\Sigma^j)^{-1}(P_1^j)^T\bigl(y^j(k) - c_{v,j}\bigr)}_{c_{x|y^j}} \\
& + \underbrace{\bigl[V_1^j (\Sigma^j)^{-1}(P_1^j)^T G_{v,j} 
\quad V_2^j M\bigr]}_{G_{x|y^j}}
\begin{bmatrix} \xi^v \\ \xi^{null} \end{bmatrix}.
\end{split}
\end{equation}
\end{small}\noindent 
Therefore, $x$ in the reachable set $\check{\mathcal{R}}_{V_1^j (\Sigma^j)^{-1} (P_1^j)^T}$ defined by $\lambda_k^j = V_1^j (\Sigma^j)^{-1} (P_1^j)^T$  is also in the set $\mathcal{Z}_{x|y^j}(k)$. \\
Based on this analysis from \eqref{eq:opt_problem} through \eqref{eq:gi_as_rm2}, we have  $$\hat{\mathcal{R}}_{k,i}^{y,\text{IN}} = \check{\mathcal{R}}_{\bar{\lambda}_k^*}\subseteq \check{\mathcal{R}}_{V_1^j (\Sigma^j)^{-1} (P_1^j)^T} \subseteq \hat{\mathcal{R}}_{k,i}^{y,\text{RM}}.$$

We next prove $\hat{\mathcal{R}}_{k,i}^{y,\text{RM}} \subseteq \hat{\mathcal{R}}_{k,i}^{y,\text{IN}}$. Consider a point $x \in \hat{\mathcal{R}}_{k,i}^{y,\text{RM}}$, represented as:
\begin{small}
\begin{equation}
x = c_{k,i} + G_{k,i}^c\xi_\text{RM}^c + G_{k,i}^b\xi_\text{RM}^b
\label{eq:rm_point_rep},
\end{equation}
\end{small}\noindent 
where $\xi_\text{RM}^c \in [-1,1]^{n_c}$, $\xi_\text{RM}^b \in \{-1,1\}^{n_b}$.
From the proposition \ref{RM}, we can find a $\xi^v  \in [-1,1]^{n_v}$ to satisfy:
\begin{small}
\begin{align}
 C^j x &= y^j - c_{v,j} - G_{v,j}\xi^v  \label{eq:RM_xi}.
\end{align}
\end{small}
With optimal weight matrices $\lambda_k^{j*}$, the IN representation are:
\begin{footnotesize}
\begin{align}
\hat{c}_{k,i}^\text{IN} &= {c}_{k,i} + \sum_{j=1}^q \lambda_k^{j*}(y^j - C^j{c}_{k,i} - c_{v,j}) \label{eq:ii_center} ,\\
\hat{G}_{k,i}^{c,\text{IN}} &= \left[(I-\sum_{j=1}^q \lambda_k^{j*} C^j){G}_{k,i}^c \quad -\lambda_k^{1*}G_{v,1} \quad \cdots \quad -\lambda_k^{q*}G_{v,q}\right] \label{eq:ii_gen_c},\\
\hat{G}_{k,i}^{b,\text{IN}} &= (I-\sum_{j=1}^q \lambda_k^{j*} C^j){G}_{k,i}^b \label{eq:ii_gen_b},\\
\label{eq:ii_con}
\hat{A}_{k,i}^{c,\text{IN}} &= \begin{bmatrix} {A}_{k,i}^c & \underbrace{0 \quad \cdots \quad 0}_{q} \end{bmatrix}, \quad \hat{A}_{k,i}^{b,\text{IN}} = {A}_{k,i}^b ,\quad  \hat{b}_{k,i}^\text{IN} = {b}_{k,i}.
\end{align}
\end{footnotesize}To prove $x \in \hat{\mathcal{R}}_{k,i}^{y,\text{IN}}$, we set:
\begin{small}
\begin{align} \label{eq:coef_c}
\xi^c &= [\xi_\text{RM}^c, \xi_v, \ldots, \xi_v]  \in [-1,1]^{n_c + q},\quad \xi^b = \xi_\text{RM}^b\in [-1,1]^{n_b},
\end{align}
\end{small}\noindent 
Checking weather \eqref{eq:coef_c} satisfies the constraints in \eqref{eq:ii_con}, we have,
\begin{small}
\begin{align}
\hat{A}_{k,i}^{c,\text{IN}} \xi^c + \hat{A}_{k,i}^{b,\text{IN}}\xi^b  &= {A}_{k,i}^c\xi_\text{RM}^c +  {A}_{k,i}^b\xi_\text{RM}^b
\end{align}
\end{small}\noindent 
According to the first row of the constraints matrices in \eqref{eq:gi_method} and  \eqref{eq:ii_con},
\begin{small}
\begin{align}
    {A}_{k,i}^c\xi_\text{RM}^c +  {A}_{k,i}^b\xi_\text{RM}^b = {b}_{k,i} = \hat{b}_{k,i}^\text{IN}.
\end{align}
\end{small}\noindent 
So, $\xi^c$ and $ \xi^b$ satisfy the constraints of the hybrid zonotope.
Substituting \eqref{eq:ii_center}-\eqref{eq:coef_c} into the IN representation:
\begin{small}
\begin{align}
&\hat{c}_{k,i}^\text{IN} + \hat{G}_{k,i}^{c,\text{IN}}\xi^c + \hat{G}_{k,i}^{b,\text{IN}}\xi^b \nonumber\\
&= {c}_{k,i} + \sum_{j=1}^q \lambda_k^{j*}(y^j - C^j{c}_{k,i} - c_{v,j}) + (I-\sum_{j=1}^q \lambda_k^{j*} C^j){G}_{k,i}^c\xi_\text{RM}^c \nonumber\\
&- \sum_{j=1}^q \lambda_k^{j*}G_{v,j}\xi_v + (I-\sum_{j=1}^q \lambda_k^{j*} C^j){G}_{k,i}^b\xi_\text{RM}^b. \label{eq:substitution}
\end{align}
\end{small}\noindent 
Putting \eqref{eq:rm_point_rep} into \eqref{eq:RM_xi}, for any $j$:
\begin{small}
\begin{equation}
y^j - C^j{c}_{k,i} - c_{v,j} = C^j({G}_{k,i}^c\xi_\text{RM}^c + {G}_{k,i}^b\xi_\text{RM}^b) - G_{v,j}\xi_v \label{eq:constraint_expand}
\end{equation}
\end{small}\noindent 
Substituting \eqref{eq:constraint_expand} into \eqref{eq:substitution} and rearranging, we get:
\begin{small}
\begin{equation}
\begin{aligned}
&\hat{c}_{k,i}^\text{IN} + \hat{G}_{k,i}^{c,\text{IN}}\xi^c + \hat{G}_{k,i}^{b,\text{IN}}\xi^b =c_{k,i} + G_{k,i}^c\xi_\text{RM}^c + G_{k,i}^b\xi_\text{RM}^b = x.
\end{aligned}
\end{equation}
\end{small}

This proves $\hat{\mathcal{R}}_{k,i}^{y,\text{RM}} \subseteq \hat{\mathcal{R}}_{k,i}^{y,\text{IN}}$. Combined with the previously proven $\hat{\mathcal{R}}_{k,i}^{y,\text{IN}} \subseteq \hat{\mathcal{R}}_{k,i}^{y,\text{RM}}$, we conclude $\hat{\mathcal{R}}_{k,i}^{y,\text{IN}} = \hat{\mathcal{R}}_{k,i}^{y,\text{RM}}$.
\end{compactproof}

Since we have shown 
$\hat{\mathcal{R}}_{k,i}^{y,\text{GI}} 
= \hat{\mathcal{R}}_{k,i}^{y,\text{RM}}$ 
and 
$\hat{\mathcal{R}}_{k,i}^{y,\text{IN}} 
= \hat{\mathcal{R}}_{k,i}^{y,\text{RM}}$, 
it follows that all three approaches yield identical sets: 
$ \hat{\mathcal{R}}_{k,i}^{y,\text{RM}} 
= \hat{\mathcal{R}}_{k,i}^{y,\text{IN}} =
\hat{\mathcal{R}}_{k,i}^{y,\text{GI}} 
$.

\section{Evaluation and Numerical Example}\label{sec:evaluation}
To validate our approach for PWA system reachability analysis with input-state data in the section \ref{state-input}, we examine a benchmark system adapted from \cite[Section~4.1.4]{bird2023hybrid}. The system dynamics are described by the following discrete-time PWA model:
\begin{small}
\begin{equation}
x[k+1]= \begin{cases}
A_1 x[k] + B_1 u[k], & \text{if } x_1 \leq 0, \\
A_2 x[k] + B_2 u[k], & \text{if } x_1 > 0,
\end{cases}\label{examplepwa}
\end{equation}
where
\begin{equation}\nonumber
A_1 = \begin{bmatrix}
0.75 & 0.25 \\
-0.25 & 0.75
\end{bmatrix}, \quad
B_1 = \begin{bmatrix}
-0.25 \\
-0.25
\end{bmatrix}
\end{equation}
\begin{equation}\nonumber
A_2 = \begin{bmatrix}
0.75 & -0.25 \\
0.25 & 0.75
\end{bmatrix}, \quad
B_2 = \begin{bmatrix}
0.25 \\
-0.25
\end{bmatrix}
\end{equation}
\end{small}We set $u[k] = \langle0,1\rangle$. Starting from an initial set defined by the zonotope:
$$
\mathcal{R}_0 = \left\langle\begin{bmatrix}
0.25 & -0.19 \\
0.19 & 0.25
\end{bmatrix}, \begin{bmatrix}
-1.51 \\
2.55
\end{bmatrix}\right\rangle.
$$

\begin{figure}[t]
    \centering
    \includegraphics[width=0.45\textwidth]{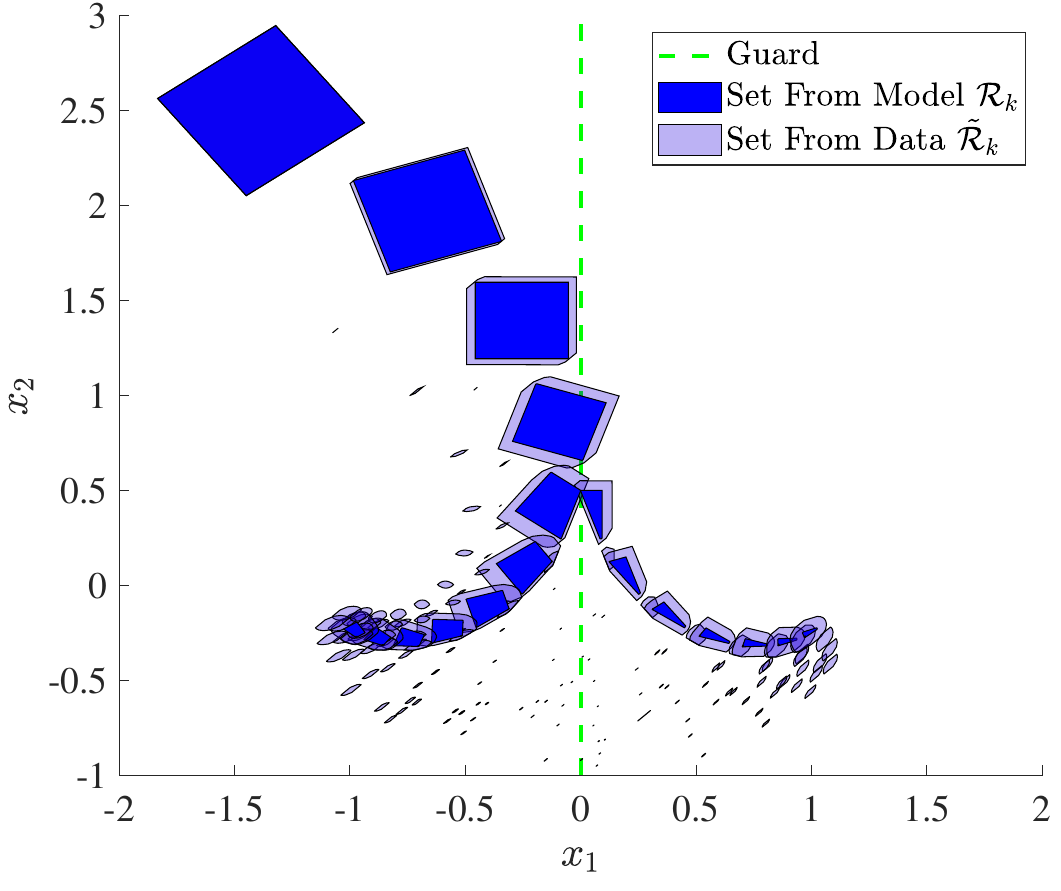}
    \caption{Reachable sets for the benchmark PWA system computed using Algorithm~\ref{alg:pwa-reachability}. The green dashed line indicates the guard condition ($x_1 = 0$) separating the two subsystems. Blue regions show analytically derived reachable sets $\mathcal{R}_k$, while light purple regions represent the data-driven approximations $\tilde{\mathcal{R}}_k$.}
    \label{fig:pwa_reachability}
\end{figure}
Figure~\ref{fig:pwa_reachability} employs Algorithm~\ref{alg:pwa-reachability} to compute and plot the state reachable set of \eqref{examplepwa}. Starting from the initial zonotope \(\mathcal{R}_0\), the reachable sets 
progress over time, switching between the two subsystems defined by 
the guard at \(x_1 = 0\). The data-driven approximations (in light purple) 
consistently encompass the model-based reachable sets (in blue), thereby 
validating the overapproximation capability of our approach. 

\begin{figure}[ht]
  \centering
   \includegraphics[width=0.45\textwidth]{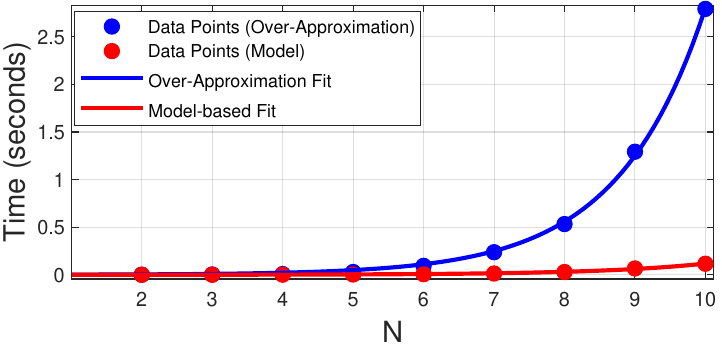}
 \caption{The experimental results show exponential growth of computation time with increasing number of steps.}
  \label{fig:pwa_example_time_computation}
\end{figure}

Figure~\ref{fig:pwa_example_time_computation} highlights the computational 
challenges when analyzing PWA systems via overapproximation methods. 
The results show a clear exponential increase in computation time with longer time horizons.
 To validate our three approaches mentioned in the section \ref{setbased} for PWA system reachability analysis with input-output data, we examine a benchmark system adapted from \cite[Section IV.
Evaluation]{alanwar2022data}:
 \begin{small}
\begin{equation}
\begin{aligned}
  A = \begin{bmatrix}
    0.9455 & -0.2426 \\
    0.2486 & 0.9455
  \end{bmatrix},
  \hspace{5mm} 
  B = \begin{bmatrix}
    0.1 \\ 0 
  \end{bmatrix}
\end{aligned}
\label{eq:system_matrices}
\end{equation}
\end{small}
with $q=3$ measurements parameterized as follows:
\begin{small}
\begin{align}\label{system_matrices2}
  &C^1 = \begin{bmatrix}
    1 & 0.4    
  \end{bmatrix}, 
 C^2 = \begin{bmatrix}
    0.9 & -1.2  
  \end{bmatrix},  
  C^3 = \begin{bmatrix}
    -0.8 & 0.2 \\ 0 & 0.7   
  \end{bmatrix}
\end{align}
\end{small}

\begin{figure}[t]
    \centering
    \includegraphics[width=0.4\textwidth]{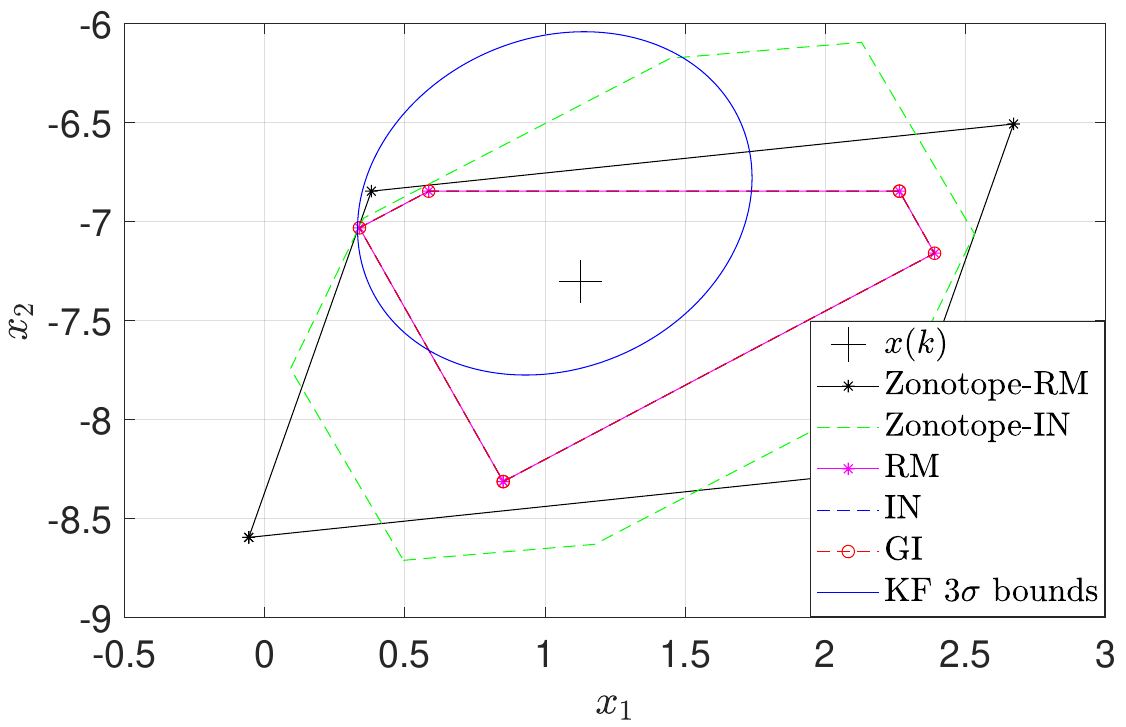}
    \caption{State-space representation of different estimation methods at time step $k$. The true state $x(k)$ is shown with a cross marker, while different estimation approaches are represented: two zonotope approximations (Zonotope-RM, Zonotope-IN), our three proposed methods (RM, IN, GI), and traditional Kalman Filter $3\sigma$ bounds.}
    \label{fig:state_space_comparison}
\end{figure}
\begin{figure}[t]
    \centering
   \includegraphics[width=0.4\textwidth]{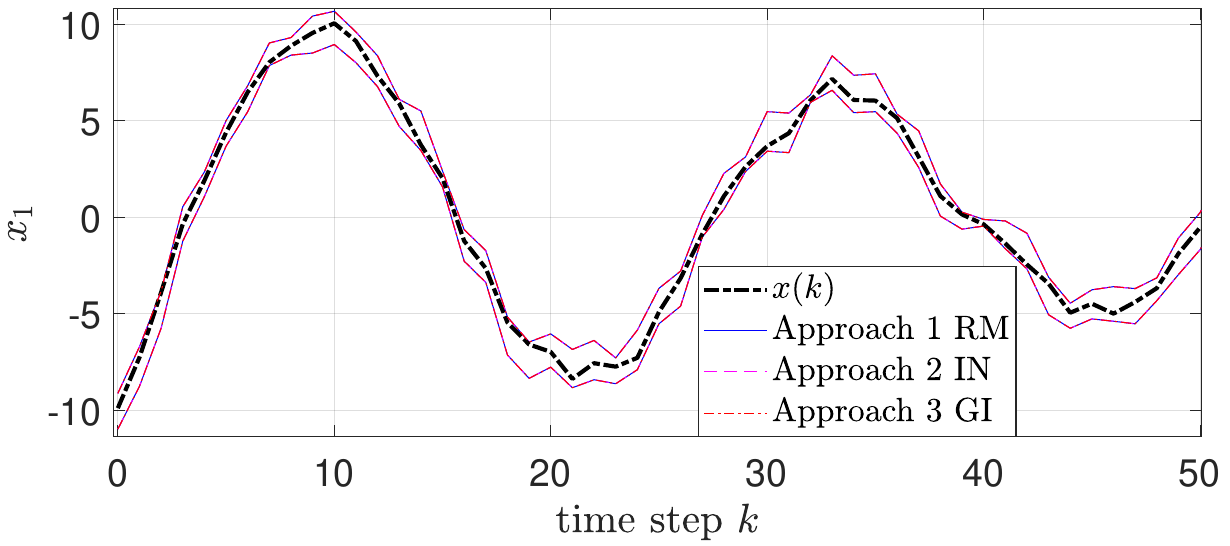}
    \caption{Set-based state estimation results for the PWA system showing the true state trajectory $x(k)$ (black dashed line) and the estimated bounds from three equivalent approaches: RM, IN, and GI.}
    \label{fig:state_estimation_comparison}
\end{figure}
\begin{figure}[t]
    \centering
    \includegraphics[width=0.45\textwidth]{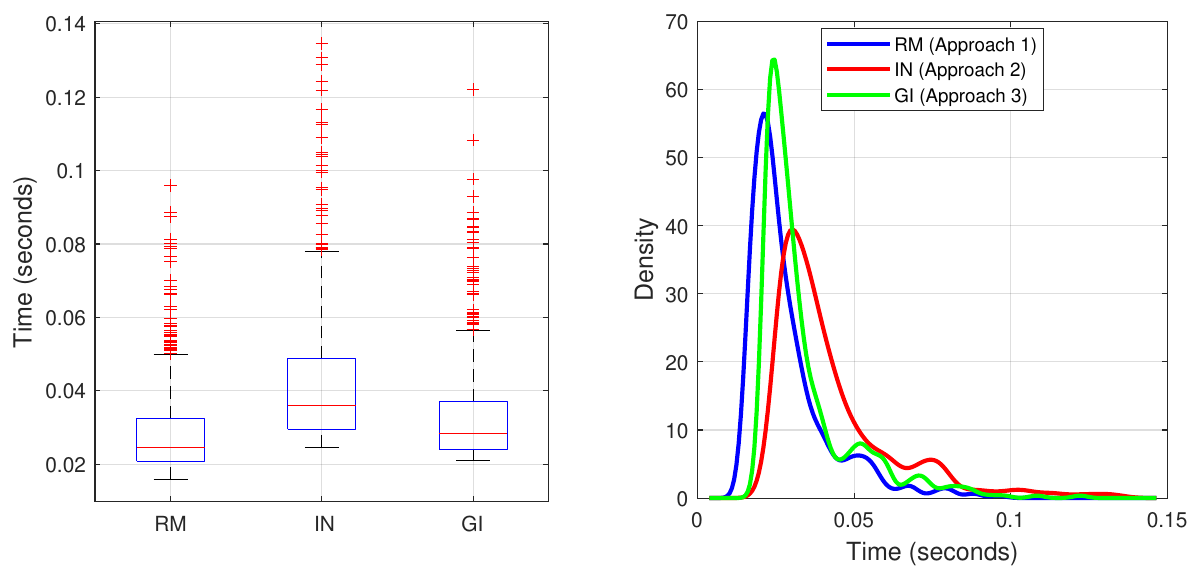}
    \caption{Computational performance comparison of the three set-based state estimation approaches. (a) Box plot showing the statistical distribution of computation times, with median values marked by central red lines and outliers as red plus signs. (b) Density plots illustrating the probability distribution of computation times for each method, demonstrating the relative efficiency of the RM, IN, and GI approaches.}
    \label{fig:computation_comparison}
\end{figure}

We evaluate the system \eqref{eq:system_matrices} and \eqref{system_matrices2} with The initial state set   
   $\mathcal{X}_0 = \langle{ [0\;\; 0]^\top, 15I_2 }\rangle$
   and the true initial state $x(0) = \begin{bmatrix} -10 & 10 \end{bmatrix}^\top$,  we get the Figures ~\ref{fig:state_space_comparison},  ~\ref{fig:state_estimation_comparison} and ,~\ref{fig:computation_comparison}.
   
Figure~\ref{fig:state_space_comparison} shows all three proposed methods (RM, IN, GI) produce identical estimation bounds, 
verifying their mathematical equivalence established by our theoretical analysis. 
These bounds differ from those of standard zonotope methods (Zonotope-RM, Zonotope-IN), 
which do not use constraint information. 
Figure~\ref{fig:state_estimation_comparison} demonstrates the equivalence of our three set-based estimation approaches. All methods (RM, IN, and GI) generate nearly identical bounds that successfully contain the true state trajectory $x(k)$ throughout the entire horizon.

Figure~\ref{fig:computation_comparison} and Table.~\ref{tab:computation_stats} compare the computational efficiency of the three estimation approaches. The box plot and density plot consistently show that the RM method achieves the lowest computation time (median: 0.029s), followed by GI (0.043s) and IN (0.034s). The table clearly demonstrates that the RM method outperforms the other approaches across all performance metrics.

\begin{table}[ht]
    \centering
    \caption{Statistical analysis of computation times (seconds)}
    \label{tab:computation_stats}
    \begin{tabular}{lcccccc}
        \hline
        Method & Mean & Median & Variance & Std Dev & Min & Max \\
        \hline
        RM & 0.029 & 0.025 & 1.73$\times$10$^{-4}$ & 0.013 & 0.016 & 0.096 \\
        IN & 0.043 & 0.036 & 3.81$\times$10$^{-4}$ & 0.020 & 0.024 & 0.135 \\
        GI & 0.034 & 0.028 & 2.35$\times$10$^{-4}$ & 0.015 & 0.021 & 0.122 \\
        \hline
    \end{tabular}
\end{table}

\section{Conclusion}\label{sec:conclusion}
This paper introduced a data-driven reachability analysis method for verifying the safety of PWA systems without prior mathematical models. By leveraging hybrid zonotopes, our approach over-approximates reachable sets using only noisy measurement data. We proved the equivalence of three set-based state estimation methods (RM, IN, and GI), demonstrating their identical estimation boundaries. Numerical experiments confirmed the effectiveness of these methods, with computational analysis highlighting exponential complexity growth and RM’s superior efficiency. This methodology opens new possibilities for safety verification in systems where accurate modeling is challenging, with significant implications for autonomous vehicles, robotic systems, and other cyber-physical applications that operate across multiple modes. Future work will focus on the coupling between the PWA system boundaries and submodel parameters, as well as the reachability analysis of other hybrid systems (e.g., mixed logical dynamical systems).

\bibliographystyle{IEEEtran} 
\bibliography{references}

@ARTICLE{amr23reachable,
  author={Alanwar, Amr and Koch, Anne and Allgöwer, Frank and Johansson, Karl Henrik},
  journal={IEEE Transactions on Automatic Control}, 
  title={Data-Driven Reachability Analysis From Noisy Data}, 
  year={2023},
  volume={68},
  number={5},
  pages={3054-3069},
  doi={10.1109/TAC.2023.3257167}}

@article{bird2023hybrid,
  title={Hybrid zonotopes: A new set representation for reachability analysis of mixed logical dynamical systems},
  author={Bird, Trevor J and Pangborn, Herschel C and Jain, Neera and Koeln, Justin P},
  journal={Automatica},
  volume={154},
  pages={111107},
  year={2023},
  publisher={Elsevier}
}

@article{heemels2001equivalence,
  title={Equivalence of hybrid dynamical models},
  author={Heemels, Wilhemus PMH and De Schutter, Bart and Bemporad, Alberto},
  journal={Automatica},
  volume={37},
  number={7},
  pages={1085--1091},
  year={2001},
  publisher={Elsevier}
}

@inproceedings{alanwar2022data,
  title={Data-driven set-based estimation using matrix zonotopes with set containment guarantees},
  author={Alanwar, Amr and Berndt, Alexander and Johansson, Karl Henrik and Sandberg, Henrik},
  booktitle={2022 European Control Conference (ECC)},
  pages={875--881},
  year={2022},
  organization={IEEE}
}

@inproceedings{asarin2000approximate,
  title={Approximate reachability analysis of piecewise-linear dynamical systems},
  author={Asarin, Eugene and Bournez, Olivier and Dang, Thao and Maler, Oded},
  booktitle={International workshop on hybrid systems: Computation and control},
  pages={20--31},
  year={2000},
  organization={Springer}
}

@inproceedings{girard2005reachability,
  title={Reachability of uncertain linear systems using zonotopes},
  author={Girard, Antoine},
  booktitle={International Workshop on Hybrid Systems: Computation and Control},
  pages={291--305},
  year={2005},
  organization={Springer}
}

@inproceedings{guernic2010zonotope,
  title={Zonotope/hyperplane intersection for hybrid systems reachability analysis},
  author={Le Guernic, Colas and Girard, Antoine},
  booktitle={International Workshop on Hybrid Systems: Computation and Control},
  pages={215--228},
  year={2010},
  organization={Springer}
}

@article{bemporad2000observability,
  title={Observability and controllability of piecewise affine and hybrid systems},
  author={Bemporad, Alberto and Ferrari-Trecate, Giancarlo and Morari, Manfred},
  journal={IEEE Transactions on Automatic Control},
  volume={45},
  number={10},
  pages={1864--1876},
  year={2000}
}

@article{paoletti2007identification,
  title={Identification of hybrid systems: a tutorial},
  author={Paoletti, Simone and Juloski, Aleksandar Lj},
  journal={European Journal of Control},
  volume={13},
  number={2-3},
  pages={242--260},
  year={2007}
}

@article{scott2016constrained,
  title={Constrained zonotopes: A new tool for set-based estimation and fault detection},
  author={Scott, Joseph K and Raimondo, Davide M and Marseglia, Giuseppe Roberto and Braatz, Richard D},
  journal={Automatica},
  volume={69},
  pages={126--136},
  year={2016},
  publisher={Elsevier}
}

@book{fiedler2006linear,
  title={Linear optimization problems with inexact data},
  author={Fiedler, Miroslav and Nedoma, Josef and Ram{\'\i}k, Jaroslav and Rohn, Jiri and Zimmermann, Karel},
  year={2006},
  publisher={Springer Science \& Business Media}
}

@article{alanwar2023distributed,
  title={Distributed set-based observers using diffusion strategies},
  author={Alanwar, Amr and Rath, Jagat Jyoti and Said, Hazem and Johansson, Karl Henrik and Althoff, Matthias},
  journal={Journal of the Franklin Institute},
  volume={360},
  number={10},
  pages={6976--6993},
  year={2023},
  publisher={Elsevier}
}

\end{document}